\input harvmac

\overfullrule=0pt
\abovedisplayskip=12pt plus 3pt minus 1pt
\belowdisplayskip=14pt plus 3pt minus 1pt
%

\def\bar{\overline}

\def\bigone{\hbox{1\kern -.23em {\rm l}}}
\def\ZZ{\hbox{\zfont Z\kern-.4emZ}}
\def\half{{\litfont {1 \over 2}}}

\font\litfont=cmr6

\font\smit=cmti8 scaled\magstep1

\lref\defo{
O.~DeWolfe, D.~Z.~Freedman and H.~Ooguri,
{\it ``Holography and defect conformal field theories,''}
Phys.\ Rev.\ D {\bf 66}, 025009 (2002)
[arXiv:hep-th/0111135].}

\lref\kp{
I.~R.~Klebanov and A.~M.~Polyakov,
{\it ``AdS dual of the critical O(N) vector model,''}
Phys.\ Lett.\ B {\bf 550}, 213 (2002)
[arXiv:hep-th/0210114].}

\lref\maldabig{
J.~M.~Maldacena, {\it ``The large N limit of superconformal field
theories and supergravity,''} Adv.\ Theor.\ Math.\ Phys.\ {\bf 2}, 231
(1998) [Int.\ J.\ Theor.\ Phys.\ {\bf 38}, 1113 (1999)]
[arXiv:hep-th/9711200].}

\lref\fvone{
E.~S.~Fradkin and M.~A.~Vasiliev,
{\it ``On The Gravitational Interaction Of Massless Higher Spin
Fields,''} Phys.\ Lett.\ B {\bf 189} (1987) 89.}

\lref\fvtwo{
E.~S.~Fradkin and M.~A.~Vasiliev,
{\it ``Cubic Interaction In Extended Theories Of Massless Higher Spin
Fields,''} Nucl.\ Phys.\ B {\bf 291}, 141 (1987).}

\lref\krone{
A.~Karch and L.~Randall,{ \it
``Locally localized gravity,''} JHEP {\bf 0105}, 008 (2001)
[arXiv:hep-th/0011156].}

\lref\leoruhal{
T.~Leonhardt and W.~Ruhl, {\it ``General graviton exchange graph for
four point functions in the AdS/CFT correspondence,''} J.\ Phys.\ A
{\bf 36}, 1159 (2003) [arXiv:hep-th/0210195].}

\lref\porrati{
L.~Girardello, M.~Porrati and A.~Zaffaroni, {\it``3-D interacting CFTs
and generalized Higgs phenomenon in higher spin theories on AdS,''}
arXiv:hep-th/0212181.}

\lref\petkouone{
A.~C.~Petkou,
{\it ``Evaluating the AdS dual of the critical O(N) vector model,''}
arXiv:hep-th/0302063.}

\lref\kirsch{
N.~R.~Constable, J.~Erdmenger, Z.~Guralnik and I.~Kirsch,
{\it ``Intersecting D3-branes and holography,''}
arXiv:hep-th/0211222.}

\lref\sezginone{
J.~Engquist, E.~Sezgin and P.~Sundell,
{\it ``On N = 1,2,4 higher spin gauge theories in four dimensions,''}
Class.\ Quant.\ Grav.\  {\bf 19}, 6175 (2002)
[arXiv:hep-th/0207101].}

\lref\sezgintwo{
J.~Engquist, E.~Sezgin and P.~Sundell,
{\it ``Superspace formulation of 4D higher spin gauge theory,''}
arXiv:hep-th/0211113.}

\lref\porratione{
M.~Porrati,
{\it ``Higgs phenomenon for 4-D gravity in anti de Sitter space,''}
JHEP {\bf 0204}, 058 (2002)
[arXiv:hep-th/0112166].}

\lref\karchone{
A.~Karch,
{\it ``Lightcone quantization of string theory duals of free field
theories,''} arXiv:hep-th/0212041.}

\lref\tseytlin{
A.~A.~Tseytlin,
{\it ``On limits of superstring in $AdS_5 \times S^5$,}
Theor.\ Math.\ Phys.\  {\bf 133}, 1376 (2002)
[Teor.\ Mat.\ Fiz.\  {\bf 133}, 69 (2002)]
[arXiv:hep-th/0201112].}

\lref\duffliusati{
M.~J.~Duff, J.~T.~Liu and H.~Sati,
{\it ``Complementarity of the Maldacena and Karch-Randall pictures,''}
arXiv:hep-th/0207003.}

\lref\klebwit{
I.~R.~Klebanov and E.~Witten,
{\it ``AdS/CFT correspondence and symmetry breaking,''}
Nucl.\ Phys.\ B {\bf 556}, 89 (1999)
[arXiv:hep-th/9905104].}

\lref\adefa{
O.~Aharony, O.~DeWolfe, D.~Z.~Freedman and A.~Karch,
{\it ``Defect Conformal Field Theory and Locally Localized Gravity,''}
hep-th/0303249.}


\lref\sumit{
S.~R.~Das and A.~Jevicki,
{\it ``Large-N Collective Fields and Holography,''}
arXiv:hep-th/0304093.}

\lref\dirac{
P.~A.~Dirac,
{\it ``A Remarkable Respresentation Of The 3+2 De Sitter Group,''}
J.\ Math.\ Phys.\  {\bf 4}, 901 (1963).}

\lref\vasilievtwo{
S.~E.~Konstein and M.~A.~Vasiliev,
{\it ``Extended Higher Spin Superalgebras And Their Massless
Representations,''} Nucl.\ Phys.\ B {\bf 331}, 475 (1990).}

\lref\vasilievthree{
M.~A.~Vasiliev,
{\it ``Higher-spin gauge theories in four, three and two
dimensions,''}  Int.\ J.\ Mod.\ Phys.\ D {\bf 5}, 763 (1996)
[arXiv:hep-th/9611024].}

\lref\vasilievfour{
M.~A.~Vasiliev,
{\it ``Higher spin gauge theories: Star-product and AdS space,''}
arXiv:hep-th/9910096.}

\lref\sungborgone{
B.~Sundborg,
{\it ``Stringy gravity, interacting tensionless strings and massless
higher  spins,''} 
Nucl.\ Phys.\ Proc.\ Suppl.\  {\bf 102}, 113 (2001)
[arXiv:hep-th/0103247].}

\lref\sezginthree{
E.~Sezgin and P.~Sundell,
{\it ``Doubletons and 5D higher spin gauge theory,''}
JHEP {\bf 0109}, 036 (2001)
[arXiv:hep-th/0105001].}

\lref\witten{
E.~Witten, talk given at J.H.~ Schwarz' 60th Birthday Conference,
Cal. Tech., Nov 2001}

\lref\sezginfour{
E.~Sezgin and P.~Sundell,
{\it ``Massless higher spins and holography,''}
Nucl.\ Phys.\ B {\bf 644}, 303 (2002)
[arXiv:hep-th/0205131].}

\lref\anselmione{
D.~Anselmi,
{\it ``Theory of higher spin tensor currents and central charges,''}
Nucl.\ Phys.\ B {\bf 541}, 323 (1999)
[arXiv:hep-th/9808004].}

\lref\anselmitwo{
D.~Anselmi,
{\it ``Higher-spin current multiplets in operator-product
expansions,''}  Class.\ Quant.\ Grav.\  {\bf 17}, 1383 (2000)
[arXiv:hep-th/9906167].}

\lref\vasilievfive{
S.~E.~Konstein, M.~A.~Vasiliev and V.~N.~Zaikin,
{\it ``Conformal higher spin currents in any dimension and AdS/CFT
correspondence,''} JHEP {\bf 0012}, 018 (2000)
[arXiv:hep-th/0010239].} 

\lref\mikhailov{
A.~Mikhailov,
{\it``Notes on higher spin symmetries,''}
arXiv:hep-th/0201019.}

\lref\dmw{
A.~Dhar, G.~Mandal and S.~R.~Wadia,
{\it ``String bits in small radius AdS and weakly coupled N = 4 super
Yang-Mills theory. I,''} 
arXiv:hep-th/0304062.}

\lref\karchtwo{
A.~Clark, A.~Karch, P.~Kovtun and D.~Yamada,
{\it ``Construction of bosonic string theory on infinitely curved
anti-de  Sitter space,''} arXiv:hep-th/0304107.}

\lref\segal{
A.~Y.~Segal,
{\it ``Conformal higher spin theory,''}
arXiv:hep-th/0207212.}

\lref\flatofronsdal{
M.~Flato and C.~Fronsdal,
{\it ``One Massless Particle Equals Two Dirac Singletons: Elementary
Particles In A Curved Space. 6,''} 
Lett.\ Math.\ Phys.\  {\bf 2}, 421 (1978).} 

\lref\heidone{
W.~Heidenreich,
{\it ``Tensor Products Of Positive Energy Representations Of SO(3,2)
And SO(4,2),''} J.\ Math.\ Phys.\  {\bf 22}, 1566 (1981).}

{\nopagenumbers
\Title{\vbox{
\hbox{hep-th/0304208}
\hbox{DAMTP-2003-36}
}}
{\vbox{
\centerline{
The Holographic Dual of a SUSY Vector Model }
\medskip
\centerline{and Tensionless Open Strings}}}
\vskip -20pt
\centerline{Nemani V. Suryanarayana}
\vskip 8pt
\medskip
\centerline{\smit DAMTP, Centre for Mathematical Sciences} 
\centerline{\smit Wilberforce Road, Cambridge, U.K} 
\centerline{\smit v.s.nemani@damtp.cam.ac.uk}
\vskip 1.0truecm
\centerline{\bf ABSTRACT}
\bigskip
We describe a supersymmetric example of the holographic duality
between 3 dimensional vector models and higher spin gauge theories in
$AdS_4$, first proposed in a bosonic context by Klebanov and
Polyakov. We argue that a particular Fradkin-Vasiliev type
supersymmetric higher spin gauge theory in $AdS_4$ with 16
supersymmetries is dual to the singlet sector of bilinear operators of
a free ${\cal N} = 4$, $SU(N)$ vector model defined on the
boundary. Starting from the duality between type IIB on $AdS_5 \times
S^5$ with a D5--brane as an $AdS_4 \times S^2$ subspace and the 4
dimensional $SU(N)$ SYM with a defect, we recover the duality between
our vector model and the higher spin gauge theory. In this case, we
propose that the higher spin gauge theory is a truncation of the open
string theory on the world volume of the D5-brane in its tensionless
string limit. We also comment on a possible Higgs mechanism in our
model.
\vfill
\Date{April 2003}
\eject}
\ftno=0

\newsec{Introduction}

Recently Klebanov and Polyakov \refs\kp\ have proposed that the
holographic dual of the singlet sector of bilinear operators of the
bosonic vector models should be certain bosonic Fradkin-Vasiliev
\refs{\fvone, \fvtwo} type
higher spin gauge theories in $AdS$ spaces. In particular, they argued
that the 3-dimensional critical $O(N)$ vector model should be dual to
the minimal higher spin gauge theory in $AdS_4$ which contains one
massless
 physical field for
each even spin $s = 0, 2, 4,
\cdots$. This duality has been motivated by the fact that the free 
$O(N)$ vector model for the real fields $\phi^a(\vec{x})$ in the
vector representation of $O(N)$:
\eqn\onvect{
S = {1 \over 2} \int d^3x\, \del^\mu \phi^a \del_\mu \phi^a}
has singlet bilinear conserved currents of the form
\eqn\oncurrents{
J_{(\mu_1 \cdots \mu_s)} = \phi^a \del_{(\mu_1} \cdots \del_{\mu_s)}
\phi^a + \cdots}
which are primaries for $s = 0, 2, 4, \cdots$. A prescription, similar
to that in the usual AdS/CFT correspondence, to find the correlation
functions of these currents from the higher spin theory has also been
proposed:
\eqn\corres{
\langle \exp \int d^3x\, h_0^{(\mu_1 \cdots \mu_s)} J_{(\mu_1 \cdots
\mu_s)} \rangle = e^{S[h_0]}}
where, as usual, $h_0$ is the boundary value of the massless spin $s$
field in the bulk and $J$ is the corresponding dual CFT operator. For
example, the massless scalar in $AdS_4$ is proposed to be dual to the
spin zero `current' $\phi^a\phi^a$. The well known infrared conformal
fixed point of the 3-dimensional vector model, in this language, is
nicely interpreted as the one obtained by deforming \onvect\ by a
`double trace' operator $(\phi^a\phi^a)^2$.

It is of interest to understand this duality in more detail. Some of
the recent developments in this direction include
\refs{\leoruhal, \porrati, \petkouone, \sumit}.

In this paper, we consider supersymmetric versions of this duality. As
a particular example, we propose that the non-minimal ${\cal N} = 4$
higher spin gauge theory \refs\sezginone\ in $AdS_4$ based on the
higher spin superalgebra $hs_0(4|4;1)$ should be holographically dual
to the singlet sector of bilinear operators of the free supersymmetric
3-dimensional ${\cal N} = 4$ theory with fields in the vector
representation of $SU(N)$.

It has been suggested in the literature that the Klebanov-Polyakov
(KP) duality should be realised as a truncation of the duality between
M-theory on $AdS_4 \times X^7$ and three dimensional CFT on the
boundary in its free field theory limit. However, it is not yet clear
if and how the KP type duality can be realised in a string theory
context. Therefore, it is interesting to study any possible
realisations of such dualities in string theory. Here, we propose an
embedding of our supersymmetric version into type IIB string theory.

We start with the well known correspondence between a 4-d `defect'
conformal field theory and the type IIB string theory on $AdS_5
\times S^5$ plus a D5--brane with its world-volume embedded as an  
$AdS_4 \times S^2$ subspace \refs{\krone, \defo}. The D5--brane acts
as a co-dimension one defect in the boundary ${\cal N} = 4$, $d=4$,
$SU(N)$ SYM.  On the 3 dimensional defect, there is matter in the
fundamental (vector) representation of the gauge group $SU(N)$. We
take a decoupling limit, the free field theory limit $\lambda = 4\pi
g^2_{YM} N \rightarrow 0$ of the defect CFT, that separates the 4-d
SYM dynamics from that of the 3-d vector model. In this limit the bulk
theory becomes that of the tensionless closed and open strings. By
taking recourse to a strong form of the AdS/dCFT duality, we argue
that the holographic dual of the singlet sector of bilinear operators
of the free vector model should be a consistent truncation of the open
string theory living on the D5--brane in the tensionless string
limit.

The paper is organised as follows. In section 2 we describe a
supersymmetric version of the KP duality. In section 3 we deal with
the subject of the string theory embedding of the duality of section
2. We also comment on how the higher string modes could become massive
through a generalised Higgs mechanism when $\lambda$ is turned on in
the defect CFT. In section 4 we give conclusions and open questions.

\newsec{A supersymmetric version of Klebanov-Polyakov duality}

In this section we describe the holographic dual of a supersymmetric
Fradkin-Vasiliev type higher spin gauge theory in $AdS_4$. We start
with a brief review of the higher spin gauge theory that we are
interested in. For a general introduction to higher spin gauge
theories of Fradkin-Vasiliev type, see \refs\vasilievfour\ and for
more details of the particular theory being considered here, see
\refs\sezginone.

\subsec{A non-minimal ${\cal N} =4$ higher spin gauge theory 
in $AdS_4$.}

Let us consider the non-minimal $U(1)_f$ extended ${\cal N} = 4$
higher spin gauge theory constructed in \refs\sezginone.\foot{It is an
extension of the minimal ${\cal N} =4$ theory of
\refs{\vasilievtwo, \vasilievthree, \vasilievfour}.} This theory is
defined in $AdS_4$ background and is based on the higher spin algebra
$hs_0(4|4;1)$. This infinite dimensional superalgebra contains
\eqn\maxsubset{
OSp(4|4) \times U(1)_f \subset hs_0(4|4;1)}
as its maximal finite dimensional subalgebra. The spectrum of massless
\foot{Recall that in $AdS_4$ the massless fields have $\Delta = s+1$ for
$s \ge {1 \over 2}$ and $\Delta = 1, 2$ for $s = 0$.} physical fields
of this theory, arranged into levels of ${\cal N}= 4$ supermultiplets,
is given in the table below
\refs\sezginone:
\smallskip
\eqn\hsspctrm{
\matrix{{{}\atop {(l,j)}}\backslash {s \atop {}} &&  && 0 &&  
{1 \over 2} && 1 && {3 \over 2} && 2 && {5
\over 2} && 3 && {7 \over 2} && 4 && \cdots \cr
&& && && && && && && && && && && \cr (-1,1) && && 6 && 4 && 1 && && &&
&& && && && \cr (0,0) && && 1+\bar 1 && 4 && 6 && 4 && 1 && && && &&
&& \cr (0,1) && && && && 1 && 4 && 6 && 4 && 1 && && && \cr (1,0) &&
&& && && && && 1 && 4 && 6 && 4 && 1 && \cr
\vdots && && && && && && && \vdots && && && &&}}
\centerline{Table(1): The higher spin spectrum based on the algebra
$hs_0(4|4;1)$.} 
\smallskip

The spectrum contains eight fields of each spin $s = 0, {1 \over 2},
1, {3 \over 2}, 2, \cdots$. At $l = -1$, we have an ${\cal N} = 4$ SYM
multiplet and at the next level a supergravity multiplet and so on.

This spectrum is identical to the massless spectrum of higher spins
obtained as a tensor product \refs{\flatofronsdal, \heidone} of two
Dirac singletons \refs\dirac\ of $OSp(4|4)$ superalgebra.
\eqn\spectrum{
{\rm Spectrum}[ hs_0(4|4;1)] \approx \Xi \otimes \Xi}
where
\eqn\diracsing{
\Xi = D({1 \over 2}, 0; {1 \over 2}, 0) \oplus D(1, {1 \over 2}; 0,
{1\over 2})}
is the Dirac singleton. The infinite dimensional UIRs of the higher
spin algebra $hs_0(4|4;1)$ are decomposed into those of $OSp(4|4)$ and
denoted by $D(\Delta, s; j_1, j_2)$. The quantum numbers are those of
the maximal compact subalgebra
\eqn\comsubalg{
SO(2)
\times SO(3) \times SU(2)_H \times SU(2)_V \subset OSp(4|4).}
One can set up the higher spin field equations \refs\sezginone\ which
we will not reproduce here. At the level $l= -1$, the vector multiplet
has six scalars which are the lowest weight components of the UIRs
$D(1, 0; {1 \over 2}, 0)$ and $D(2, 0; 0, {1 \over 2})$. Similarly,
the scalars at the next level are the lowest weight components of
$D(1, 0; 0, 0)$ and $D(2, 0; 0, 0)$. The rest of the fields have
$\Delta = s+1$ and all the 6s in the table above are made out of
$(1,0) \oplus (0,1)$ and all the 4s are made out of $({1\over 2}, 0)
\oplus (0, {1\over 2})$ of $SU(2)_H
\times SU(2)_V$. The whole spectrum is neutral under $U(1)_f$ by
construction. Next, we construct a dual vector model.

\subsec{The dual vector model}

We look for a $d=3$, $SU(N)$ vector model with ${\cal N}= 4$
supersymmetry with the following properties. It should admit conserved
currents ${\cal J}^{(i)}_s (\vec{x})$ (with $i= 1,2, \cdots 8$ and $s
= 0, {1\over 2}, 1, \cdots$) belonging to the singlet sector of
$SU(N)$ and are bilinear in the fundamental fields. It should admit
$OSp(4|4) \times U(1)_f$ symmetry group. The conserved currents ${\cal
J}^{(i)}_s (\vec{x})$ should have the right quantum numbers of
$OSp(4|4)$ and neutral under $U(1)_f$ with multiplicities as the
corresponding higher spin fields listed in the previous subsection.

We have seen that the bulk field content can be obtained as a tensor
product of two Dirac singletons \diracsing. Therefore, it is natural
to choose the field content of the boundary theory to have the same
quantum numbers as the lowest weight states of the singleton. That is,
a doublet of complex scalar fields $q^m(\vec{x})$, $m = 1,2$, and a
doublet of spinors $\psi^i(\vec{x})$, $i = 1, 2$, both in the
fundamental representation of $SU(N)$ with their conjugates
transforming in the anti-fundamental. The field content is summarised
below:
\smallskip
\eqn\ffields{
\matrix{{\rm Field} && \Delta && {\rm Spin}\, (s) && SU(2)_H && 
SU(2)_V && SU(N) && U(1)_f \cr
q^m && {1 \over 2} && 0 && {1\over 2} && 0 && N && 1 \cr
\psi^i && 1 && {1 \over 2} && 0 && {1 \over 2} && N && 1}}
where $m$ and $i$ are the double indices of $SU(2)_H$ and $SU(2)_V$
respectively. A free supersymmetric action with this field content can
be written down:
\eqn\bcft{
S_{boundary} = \int d^3x \,[(\del^k q^m)^\dagger \del_k q^m - i
\bar\psi^i \rho^k \del_k \psi^i]}
where $\rho_k$ are the Dirac matrices in $d=3$. This free theory is
superconformal and admits $OSp(4|4)\times U(1)_f$ as its symmetry
group. Eight of its supercharges correspond to the ordinary
supersymmetries and the remaining eight generate superconformal
supersymmetries.

Following \refs\kp\ let us construct a set of conserved currents in
this theory which are singlet bilinear operators and associate them
with the corresponding bulk fields listed in the previous
subsection.\foot{For details on the subject of higher spin conserved
currents in conformal field theories, see Refs. \refs{\anselmione, 
\anselmitwo, \vasilievfour, \vasilievfive, \mikhailov}.} First, there 
are eight scalar ($s=0$) operators:
\eqn\scalars{
{\bar q}^m \sigma^I_{mn} q^n, ~~~ {\bar \psi}^i \sigma^{A}_{ij}
\psi^j, ~~~ {\bar q}^m q^m~~ {\rm and}~~ {\bar \psi}^i \psi^i}
where $I$ is a triplet index of $SU(2)_H$ and $A$ is a triplet index
of $SU(2)_V$. $\sigma^I$ and $\sigma^A$ denote the Pauli matrices. The
three scalars ${\bar q}^m \sigma^I_{mn} q^n$ ($I = 1, 2, 3$) have
$\Delta = 1$ and have the correct quantum numbers to be dual to three
of the scalars at level $l = -1$ in table (1). Similarly,
operators ${\bar \psi}^i \sigma^{A}_{ij}
\psi^j$ correspond to the remaining three scalars in the ${\cal N} =4$
SYM multiplet of the table (1). The remaining two operators ${\bar
q}^m q^m$ and ${\bar \psi}^i \psi^i$ correspond to the scalars in the
$(l, j) = (0, 0)$ multiplet of table (1).

At $s = {1 \over 2}$, we have 
\eqn\spinors{
\psi^{\star i} q^m ~~~ \hbox{and} ~~~ {\bar q}^m \psi^i}
which are again eight in number and have the right quantum numbers to
be associated with the eight spin half fields appearing in table
(1). Let us now turn to the vectors. We find the following
eight conserved currents with $s=1$:
\eqn\vectors{
\bar q^m {\,\del^k}{\!\!\!\!\!\!\!{}^{{}^\leftrightarrow}}~ q^m, ~~~
{\bar \psi}^i \rho^k \psi^i, ~~~ \bar q^m
\sigma^I_{mn} {\,\del^k}{\!\!\!\!\!\!\!{}^{{}^\leftrightarrow}}~ q^n
~~ {\rm and} ~~ {\bar \psi}^i \sigma^A_{ij}\rho^k \psi^j.}
We associate these currents with the eight vector fields in table (1)
with the appropriate quantum numbers.

In general, at any given integer spin $s \ge 0$ we can construct eight
conserved currents of the form
\eqn\spinsone{
{\cal J}^\alpha_s (\vec{x}) = \bar q^m
\sigma^\alpha_{mn}{\,\del^{(k_1}{\!\!\!\!\!\!\!\!\!\!\!
{}^{{}^\leftrightarrow}}}~~
{\,\del^{k_2}{\!\!\!\!\!\!\!\!\!{}^{{}^\leftrightarrow}}}~ 
\cdots {\,\del^{k_s)}{\!\!\!\!\!\!\!\!\!\!\!{}^{{}^\leftrightarrow}}}~
~~q^n + \cdots,}
\eqn\spinstwo{
{\cal J}^\beta_s (\vec{x}) = {\bar \psi}^i
\sigma^{\beta}_{ij}\rho^{(k_1}
{\,\del^{k_2}{\!\!\!\!\!\!\!\!\!{}^{{}^\leftrightarrow}}}~ 
\cdots {\,\del^{k_s)}{\!\!\!\!\!\!\!\!\!\!\!{}^{{}^\leftrightarrow}}}~~
~\psi^j + \cdots}
where $\alpha, \beta = 0, 1, 2, 3$ with $\sigma^0 = I_{2 \times
2}$. Similarly, there are eight conserved currents for each
half-integral spin:
\eqn\spinthree{
{\cal J}_s^{im} = \psi^{\star i} {\,\del^{(k_1}{\!\!\!\!\!\!\!\!\!\!\!
{}^{{}^\leftrightarrow}}}~~
{\,\del^{k_2}{\!\!\!\!\!\!\!\!\!{}^{{}^\leftrightarrow}}}~ 
\cdots {\,\del^{k_s)}{\!\!\!\!\!\!\!\!\!\!\!{}^{{}^\leftrightarrow}}}~
~~ q^m + \cdots,}
\eqn\spinfour{
{\bar{\cal J}}_s^{im} = {\bar q}^m {\,\del^{(k_1}{\!\!\!\!\!\!\!\!\!\!\!
{}^{{}^\leftrightarrow}}}~~
{\,\del^{k_2}{\!\!\!\!\!\!\!\!\!{}^{{}^\leftrightarrow}}}~ 
\cdots {\,\del^{k_s)}{\!\!\!\!\!\!\!\!\!\!\!{}^{{}^\leftrightarrow}}}~
~~ \psi^i + \cdots.}
Notice that these currents have the right quantum numbers of
$OSp(4|4)$ and are neutral under $U(1)_f$ as required. For more
complete expressions of these currents please see the
appendix.\foot{Notice that all the higher spin fields can be packaged
into the following set of bi-local collective fields in the sense of
refs\sumit: ${\bar q}^m (\vec{x}) \sigma^\alpha_{mn} q^n(\vec{y})$,
${\bar \psi}^i (\vec{x}) \sigma^\beta_{ij} \rho^\mu \psi^j (\vec{y})$,
$\psi^{\star i}(\vec{x}) q^m (\vec{y})$ and ${\bar q}^m (\vec{x})
\psi^i (\vec{y})$ with ${\rho^\mu = \{I_{2\times 2}, \rho^k : k = 0,
1, 2 }\}$.}

Thus we conclude that our free ${\cal N} = 4$ $SU(N)$ vector model
\bcft\ admits the right set of bilinear operators in its 
singlet sector that are in one to one correspondence with the fields
of the higher spin gauge theory of the previous subsection. Of course,
one needs to arrange these currents into supermultiplets which may
involve taking various linear combinations of the basic currents
listed above. 

One can easily generalise the precription Eq.\corres\ of \refs\kp\ to
evaluate the boundary correlators using the bulk higher spin theory to
our supersymmetric context. The procedure followed here can clearly be
generalised to construct the dual of any given higher spin gauge
theory whose field content is contained in the tensor product of its
Dirac singletons.

Let us now turn to finding an embedding of the model studied here into
a string theory context.

\newsec{Vector models and tensionless open strings}

It has been conjectured in the literature that there exists an
unbroken symmetric phase of M-theory on $AdS_4 \times X^7$ backgrounds
dual to an $SU(N)$ invariant free singleton theory on the boundary. To
incorporate the Klebanov-Polyakov type dualities into this set up it
has been suggested (see, for instance, \refs\sezgintwo\ and references
there in) that a massless sector of M-theory on $AdS_4 \times X^7$
should be dual to the singlet sector of $O(N^2 -1)$ vector model. In
what follows, we argue that a more natural set up for realising a
Klebanov-Polyakov type duality is type II string theories in the
presence of D-branes.\foot{For discussions on higher spin theories
from type II closed strings, see
\refs{\sungborgone,
\sezginthree, \vasilievfive, \witten, \sezginfour}.} We concentrate 
on the particular example discussed in the previous section.

As was outlined in the introduction, the set-up of interest is a
Karch-Randall compactification \refs\krone\ of type IIB string
theory. That is, we embed a D5--brane into the the $AdS_5
\times S^5$ background as an $AdS_4 \times S^2$ subspace. This theory
is expected to be dual to a `defect' CFT on the boundary generalising
the usual AdS/CFT correspondence \refs\maldabig. The details
of this duality along with a perturbative definition of the the dual
field theory have been worked out in
\refs\defo (See also \refs\kirsch). We are going to use this duality
in an essential way. So let us first review briefly a few relevant
details of this duality.

\subsec{The bulk theory}

The bulk side of this set up contains type IIB closed strings
propagating in the maximally supersymmetric $AdS_5 \times S^5$
background along with the D5--brane open strings propagating in the
$AdS_4 \times S^2$ subspace. This system can be obtained by starting
with $N$ D3--branes along $(x^1, x^2, x^6)$ directions plus one
D5--brane along $(x^1, x^2, x^3, x^4, x^5)$ directions (with both D3
and D5--branes sitting at the same point in $(x^7, x^8, x^9)$
directions) and then taking the near horizon limit of the
D3--branes. Before taking the large $N$ limit, one has closed strings
along with open strings in the Chan-Paton sectors 3-3, 3-5 and
5-5. The 3-3 sector fields are in the adjoint representation of
$U(N)$, 3-5 strings in the bifundamental representation of $U(N)
\times U(1)$ and the 5-5 strings contain a $U(1)$ gauge field. Taking
the near horizon limit separates the 3-3 and 3-5 strings from the 5-5
and closed strings.

The isometry group of the resultant background is 
\eqn\isometries{
SO(2,3) \times
SO(3) \times SO(3) \sim Sp(4) \times SU(2)_H \times SU(2)_V \subset
SO(2,4) \times SO(6).}
The latter is the isometry group of $AdS_5 \times S^5$ without the
D5--brane. The presence of the D5--brane breaks half of the 32
supersymmetries of $AdS_5 \times S^5$ and we have
\eqn\supgroups{
OSp(4|4) \subset SU(2,2|4)}
as the unbroken supergroup. There is also a $U(1)_f$ gauge field on
the D5--brane. The low energy description of this system is studied in
\refs\defo\ and the modes are classified according to the
representations of $OSp(4|4)$. Since the supergravity and the DBI
modes are expected to survive the tensionless string limit, they will
be useful for our purposes later on. Let us review the relevant open
string results found in \refs\defo\ here. For more details see section
3.2 of Ref.\refs\defo.

There are three sectors for the light 5-5 open string modes: $(i)$ The
angular fluctuation ($\psi$) of the D5--brane on $S^5$, $(ii)$ The
gauge field fluctuation ($b_\mu$) of the D5--brane along the $AdS_4$
directions and $(iii)$ two coupled sectors $(b + z)^{(\pm)}$ of gauge
field components along $S^2$ and the remaining transverse
fluctuations. Below, we summarise the spectrum of these fluctuations in
terms of their $AdS_4$ energies (the conformal dimensions of the dual
operators):
\eqn\fivefive{
\eqalign{
(i)~~~& \Delta_+ = 2 + l, ~~~~ \Delta_- = 1 - l, \cr
(ii)~~~& \Delta = 2 + l, \cr
(iii)~~~& \Delta^{(+)}_\pm = {3 \over 2} \pm {1 \over 2} (2l + 5),
~~ \Delta^{(-)}_\pm = {3 \over 2} \pm {1\over 2} |2l - 3|.}}
Here $l$ labels the spherical harmonics over the $S^2$ that the
D5--brane wraps, whose symmetry we denote by $SU(2)_H$. The masses of
these modes are related to their conformal dimensions by the usual
relations. Notice that $\Delta_-$ is allowed only for $l=0$,
$\Delta^{(+)} _-$ is not possible at all and $\Delta^{(-)}_-$ is
allowed only for $l = 1,2$ by unitarity.\foot{Recall that the
unitarity in $AdS_4$ requires the $\Delta \ge s + {1 \over 2}$ for $s
= 0, {1 \over 2}$ and $\Delta \ge s + 1$ for $s = 1, {3\over 2}, 2,
\cdots$.} Under the $SU(2)_V$ symmetry the vector $b_\mu$ is a
singlet, the scalars $\psi$ and $(b + z)^{(-)}$ are a triplet and a
singlet respectively.

Further, for $l=0$, the sectors $(i)$ and $(ii)$ contain three
massless scalars (a triplet of $SU(2)_V$) and a massless vector
respectively. Similarly, the $\Delta_{\pm}^{(-)}$ sector contains
another three massless scalars at $l=1$ (a triplet of
$SU(2)_H$). There are also closed string supergravity modes which we
do not describe here. Beyond the low energy limit, there will be
higher stringy modes as well.

\subsec{The boundary theory}

Now, let us briefly review the relevant information about the boundary
theory dual to the system described above. The intersection of the
D5--brane on the boundary of $AdS_5 \times S^5$ is a co-dimension one
defect in the 4-d CFT. The defect carries, on its world volume, a
$3$-dim vector model with one hypermultiplet in the fundamental
representation of the gauge group. 

Therefore, the boundary theory is a `defect' CFT (or simply a
dCFT). This is expected to be a superconformal field theory with 16
supersymmetries. The symmetry group is again the superconformal group
$OSp(4|4)$ of which the $SO(2,3)$ is the conformal group and the
$SO(4) = SU(2)_H \times SU(2)_V$ is the global R-symmetry. There is
also a $U(1)_f$ global symmetry under which the defect fields are
charged.\foot{This is the remnant of the fact that the defect fields
belonged to the sector of strings between D3 branes and a
D5-brane. The $SU(N)$ is a gauge symmetry but $U(1)_f$ is only a
global symmetry after taking the near horizon limit.} A lagrangian
formulation of this theory is described in
\refs\defo. Here we will be content with describing the field content
of the theory and refer the reader to the original work for more
details. The 4-d ambient CFT (which we refer to as $a\rm FT$) fields
are the vector $A_k, A_6$ for $k= 0, 1, 2$, the scalars $X^I_H$ and
$X^A_V$ where $I = 3, 4, 5$ and $A = 7, 8, 9$ and the Majorana spinors
$\lambda_{im}$ where $i$ and $m$ are the doublet indices of $SU(2)_V$
and $SU(2)_H$ respectively. All these fields are in the adjoint
representation of the gauge group $SU(N)$. Further, we have the fields
on the defect: the 3--d scalars $q^m$ and the fermions $\psi^i$ which
are doublets of $SU(2)_H$ and $SU(2)_V$ respectively. Both these
fields are in the fundamental representation of the gauge group
$SU(N)$. Notice that the quantum numbers of these fields are the same
as the ones given in Eq.\ffields.

The operators dual to the light open string modes described in the
previous subsection are to be identified with gauge invariant single
trace operators made of the fundamental fields of the field theory on
the defect (which we refer to as $d\rm FT$) and the adjoint fields of
$a\rm FT$ restricted to the defect.

A dictionary of such identifications can be found in section 5 of
Ref.\refs\defo\ which we reproduce below for later reference.
\eqn\sugramodes{
\matrix{{\rm Mode} && \Delta && SU(2)_H && SU(2)_V && {\rm Operator}\cr
\noalign{\medskip}
b_\mu && l+2 && l \ge 0 && 0 && i\bar q^m
{\,D^k}{\!\!\!\!\!\!\!{}^{{}^\leftrightarrow}}~ q^m + {\bar \psi}^i
\rho^k \psi^i \cr
\noalign{\smallskip}
\psi & & l+2, \, 1-l && l \ge 0 && 1 && \bar\psi_i \sigma^A_{ij}
\psi_j
+ 2 {\bar q}^m X^{A}_V q^m \cr
\noalign{\smallskip}
(b + z)^{(-)} && l, \, 3-l && l \ge 1 && 0 && \bar q^m \sigma^I_{mn}
q^n \cr}}
where, in the last column, we indicated the dual operators
corresponding to the lowest $l$ modes which form a ${\cal N} = 4$,
$d=4$, $U(1)$ SYM multiplet as expected from the supersymmetry
counting.\foot{The mode $(b+z)^{(+)}$ is not listed above as it is
always massive and drops out of our story.} The dual operators for
higher $l$ modes are also proposed in \refs\defo. For example:
\eqn\highlops{
C_{l+1}^{I_0 \cdots I_l} = {\bar q}^m \sigma^{(I_0}_{mn} X_H^{I_1} \cdots
X_H^{I_l)} q^n - \rm traces}
gives the operator dual to a mode with the $SU(2)_H$ quantum number
$l+1$ in the last row of \sugramodes.

The boundary theory is argued to be an exact CFT
\refs{\defo, \kirsch}. Therefore, one expects that the 
$AdS_4 \times S^2$ embedding of the D5--brane is a solution to the
string equations of motion. We are going to assume that there exists
an exact world sheet boundary conformal field theory of this D5--brane
in $AdS_5 \times S^5$ background. We will also assume that the duality
is valid in its strongest form for all values of the 't Hooft coupling
$\lambda$ and $N$ and treat the field theory as a definition of the
dual string theory whenever necessary.

Having identified a type IIB string theory context in which the
3-dimensional field content \ffields\ of the supersymmetric vector
model of section 2 is realised, we next aim to carefully extract the
vector model from the boudary theory and obtain its dual bulk theory
using the AdS/dCFT correspondence reviewed above.

\subsec{Super KP from AdS/dCFT}

To summarise, on the bulk side of the duality there is the theory of
type IIB closed strings in $AdS_5 \times S^5$ coupled to the open
strings on a D5--brane embedded as an $AdS_4 \times S^2$ subspace. On
the boundary, we have the defect CFT. Holography tells us that these
two systems describe the same physics. Therefore, we write
\eqn\ideaone{
\eqalign{
& {\cal L}^{(Bulk)}_{IIB~Closed~Strings} + {\cal L}^{(Bulk)}_{D5-Open~
Strings} + {\cal L}^{(Bulk)}_{Int.} \cr 
&~~~~~~~~~~ \approx {\cal L}^{(Bdy)}_{d=4,\, {\cal
N} = 4,\, SU(N),\, SYM} + {\cal L}^{(Bdy)}_{d=3,\, {\cal N} = 4,
\,SU(N)\, Vector~Model} + {\cal L}^{(Bdy)}_{Int.} }}
where we have denoted the theories by their Lagrangian densities
schematically. On the right hand side, the first two terms represent
$a\rm FT$ and $d\rm FT$ respectively and the third denotes the
interactions between them.

Since we are interested in singling out the vector model, we first
need to decouple both $a\rm FT$ and $d\rm FT$. Therefore, we seek a
limit of the boundary theory in which the term ${\cal
L}^{(Bdy)}_{Int.}$ can be dropped. This indeed happens in the limit
$\lambda \rightarrow 0$ where $\lambda = 4\pi g^2_{YM}N$ is the 't Hooft
coupling on the boundary\foot{For this limit we keep $N$ to be large
but finite and take $g^2_{YM} \rightarrow 0$.}  and $\lambda = {R^4
\over \alpha'^2}$ in the bulk. In this limit, both ${\cal
L}^{(Bdy)}_{4-d,\, {\cal N} = 4,\, SU(N),\, SYM}$ and ${\cal
L}^{(Bdy)}_{3-d,\, {\cal N} = 4,
\,SU(N)\, Vector~Model}$ become free field theories and the 
interaction lagrangian ${\cal L}^{(Bdy)}_{Int.}$ goes to zero. On the
bulk side, the limit $\lambda \rightarrow 0$ is the tensionless string
limit.\foot{In a strict large $N$ limit one expects that ${\cal
L}^{(Bulk)}_{Int.}$ also should go to zero. But we will be interested
in keeping higher orders in $1/N$.} Therefore, in this limit, one
expects that the whole string spectrum to become massless with
infinitely many states at each spin.

Now that we have decoupled the $a\rm FT$ and $d\rm FT$, we can in
principle, extract the holographic dual of the vector model of $d\rm
FT$ alone. But unfortunately we do not have much handle over the bulk
side of the story. Therefore, we take recourse to indirect arguments
based on the assumption that the strong form of the AdS/dCFT is true.

First we restrict ourselves to single trace operators of the boundary
theory. The calculable quantities of interest in the CFT are the
correlation functions of gauge invariant operators. In the limit
$\lambda \rightarrow 0$, these operators are simply the singlets of
$SU(N)$. There are essentially three classes of these singlet
operators. Denoting an arbitrary adjoint field of $a\rm FT$ by ${\cal
A}$ and a fundamental field of $d\rm FT$ by ${\cal F}$ these
operators, schematically, are of the type:
\eqn\typesofops{
(a) ~~~ {\rm Tr} ({\cal A} \cdots {\cal A}), ~~
~~ (b)~~~{\bar {\cal F}} {\cal A} \cdots {\cal A} {\cal F}, ~~
~~(c)~~~{\bar {\cal F} {\cal F}}.}
The operators of type $(a)$ correspond to the closed string
states. The operators of type $(b)$ and $(c)$ (in general,
combinations of those, like in Eq.(3.4)) correspond to the open string
states.\foot{There is, in general, operator mixing with multitrace
operators which is crucial at higher orders in perturbation
theory. Here, we mean the operators which reduce to the ones listed in
\typesofops\ at the lowest order in the perturbation theory.}

We are interested in the vector model and therefore in the correlation
functions involving type $(c)$ operators alone. Hence we need to
`subtract' the $a\rm FT$ from the boundary theory. At the level of
correlation functions, this amounts to setting all the operators
belonging to type $(a)$ and $(b)$ to zero which is achieved by simply
setting the fields of the $a\rm FT$ to zero. This is consistent
because we have decoupled $a\rm FT$ and $d\rm FT$ completely. 

By `subtracting' the $a\rm FT$ from the boundary theory, we are left
with a much reduced set of operators (those of type $(c)$) on the
boundary theory. Let us try to infer what this truncation of the set
of operators of the boundary theory means for the bulk theory. Using
the strong form of the AdS/dCFT correspondence, we conclude that
setting type $(a)$ operators to zero amounts to setting all the dual
closed string states on $AdS_5
\times S^5$ to zero. Similarly, setting type $(b)$ operators to zero
corresponds to truncating the spectrum of tensionless open string
states consistently.

More specifically, out of the spectrum of the light open string modes
discussed in section (3.1) (all of which survive the tensionless
string limit) we will be left with only the lowest $l$ modes after the
truncation. This is because the operators dual to any higher $l$ mode
contains powers of adjoint fields of $a\rm FT$ (like in
\highlops) which we set
to zero as they belong to type $(b)$ of \typesofops. Thus, we get rid
of all the higher KK modes of these light states in this
truncation. About the higher string modes we again go back to the dual
theory. For this we simply note that the free field theory that we
obtain in the limit is exactly the one we considered in section
2. There, we constructed all the operators that could be dual to the
string states that we are after. Therefore, we expect that the
truncated spectrum of the tensionless open strings on $AdS_4
\times S^2$ should be identical to the spectrum of the higher spin 
gauge theory on $AdS_4$ discussed in section 2.

In other words, the higher spin gauge theory considered in section 2
is precisely the consistent truncation of the tensionless open string
theory. The truncation is achieved by subtracting the $d=4$ $a\rm FT$
from the AdS/dCFT correspondence that we started with. The symmetry
group $OSp(4|4)$ is now simply the superconformal group of the 3-d
vector model and the $U(1)_f$ is now associated with the $U(1)$ gauge
field on the D5--brane. The neutrality of the higher spin spetrum of
section 2 under $U(1)_f$ is simply the reflection of the fact that the
open string states on a single D-brane are neutral under its $U(1)$
gauge field. However, if we had more than one coincident D5-branes,
the corresponding higher spin spectrum would have been charged under
the the non-abelian gauge field on the branes.

It may be puzzling that the gravity theory expected for holography is
now actually part of a tensionless open string theory living on a
D5-brane. However, one expects a graviton (a massless spin two
particle) in the open string spectrum in the tensionless string
limit. One way of arguing for this is to look at the conserved
energy-momentum tensor of the dual boundary CFT \refs\defo :
\eqn\emtensor{
T_{\mu\nu}(\vec{y}, x) = {\cal T}_{\mu\nu} (\vec{y}, x) + \delta(x)
\delta_\mu^k \delta_\nu^l t_{kl}(\vec{y})}
where, $\vec{y}$ and $x$ are the coordinates along and transverse to
the defect. Neither ${\cal T}_{\mu\nu}$ nor $t_{kl}$ is conserved by
itself in the full theory. In the free CFT limit, however, they are
both conserved individually. Since in this limit $t_{kl}$ is quadratic
in the fields of the $d\rm FT$, it corresponds to an open string
mode. Hence, we expect that one of the stringy modes of the 5-5 sector
to become massless exactly in this limit and play the role of the
graviton dual to this conserved energy momentum tensor.\foot{This was
also noticed recently in \refs\adefa.} Notice that there are eight
spin 2 fields in the higher spin spectrum of table (1). But we expect
that the actual graviton dual to the 3-d energy momentum tensor is a
combination of the spin 2 states in the second and fourth rows of
table (1) as the others do not have the same global quantum numbers
$(j_1, j_2)$ as $t_{kl}$.

Before concluding this section, we comment on the possibility of
understanding how the higher open string states of the higher spin
theory could attain masses once we go away from the decoupling limit
discussed above.

\subsec{The open string graviton and the higgs mechanism}

For consistency we expect that, when $\lambda$ is turned on, all the
fields except for the massless ${\cal N} = 4$ SYM multiplet (at level
$l=-1$ in table (1)) acquire masses proportional to the string
tension. From the boundary theory point of view, this means that the
operators dual to these states develop anomalous dimensions
\eqn\estimate{
\Delta - s - 1 \sim {\cal O} (\lambda)}
In the full interacting dCFT, one can see this happening in
perturbation theory \refs{\defo, \adefa}. From the bulk theory point
of view, once we turn on the tension of the strings, we can no longer
neglect interactions with closed string modes (and the rest of the
open string modes which we set to zero by the consistent truncation
explained above). But it may be possible to argue that the extra
massless modes acquire mass for nonzero $\lambda$ in the higher spin
theory in $AdS_4$. A related issue has been considered in the
literature \refs\porratione\ (see also
\refs\duffliusati) for $s = 2$.\foot{There is, however,
an essential difference between \refs\porratione\ and our case. In
\refs\porratione, Higgs mechanism for the the bulk (closed
string) graviton was considered ({\it \`a la} Karch-Randall
brane-world \refs\krone). We are interested in the open string
graviton.}  In Ref.\refs\porratione\ it has been argued that depending
on the boundary conditions set for the massless scalar, the graviton
goes through a Higgs mechanism to acquire mass. By simply noting that
the required mixing of boundary conditions occurs precisely when the
defect interacts with the ambient CFT, we can take over the results of
\refs\porratione\ into our context.

Massless UIRs of $AdS_4$ group with spin $s > 0$ have $\Delta = s + 1$
and massive ones have $\Delta > s + 1$. A massive UIR $D(\Delta,\,
s;\, j_1,\, j_2)$ becomes reducible in the massless limit $\Delta
\rightarrow s + 1$. For example, for a spin $s$ massive UIR in $AdS_4$
denoted by $D(\Delta, s)$ one has the decomposition
\eqn\decomp{
D(\Delta, s) \rightarrow D(s + 1, s) \oplus D(s + 2, s - 1)}
in the limit $\Delta \rightarrow s + 1$. The first piece on the r.h.s
is massless and the second is a massive representation. This implies
that in $AdS_4$ a spin $s$ field must eat a massive particle of spin
$s-1$ to attain mass. Generalising this result to our case, the
decomposition of a massive representation $D(\Delta, s, j_1, j_2)$ in
the limit of $\Delta
\rightarrow s + 1$ may be written as :
\eqn\superdecomp{
D(\Delta, s, j_1, j_2) \rightarrow D(s+1, s, j_1, j_2) \oplus D(s+2, s
- 1, j_1, j_2).}
This gives the quantum numbers required for the Higgs fields. One
should be able to find these representations in the tensor product of
the scalar field representations of table (1) as explained in
\refs\porrati. We postpone a more detailed analysis to the future.

\newsec{Conclusion}

We have provided a supersymmetric example of the holographic duality
between vector models and higher spin gauge theories first considered
by Klebanov and Polyakov \refs\kp\ in a bosonic context. Our bulk
theory is a particular non-minimal ${\cal N} = 4$ higher spin gauge
theory in $AdS_4$ and based on the algebra $hs_0(4|4;1) \supset
OSp(4|4) \times U(1)_f$. It contains eight physical fields at each
spin $s = 0, {1 \over 2}, 1, {3 \over 2}, 2, \cdots$. The holographic
boundary theory is a 3-dimensional ${\cal N} = 4$ $SU(N)$ free vector
model. Since our method is based on constructing the field theory
using the singleton representations of the higher spin algebra, it
should help constructing other examples where the higher spin field
content is contained in the tensor product of the corresponding Dirac
singleton.

We have suggested that the model studied here can be embedded into
type IIB string theory. We have argued that the bulk theory should be
a truncation of the open string theory on a D5 brane, embedded
as $AdS_4 \times S^2$ subspace of $AdS_5 \times S^5$ background of
type IIB, in its tensionless string limit. The truncation is
understood from the boundary theory point of view simply as the
`subtraction' of the free $a\rm FT$.

It is amusing that a large $N$ CFT is dual to an open string theory as
opposed to a closed string theory. The role of the graviton is
expected to be played by a massive open string mode which becomes
massless in the tensionless string limit. One could consider variants
of the model described in this paper. For example, taking an
orientifold of the model considered here, where one replaces $AdS_5
\times S^5$ by $AdS_5 \times RP^5$, should result in an $SO(N)$
version of our duality. One can use the notion of consistent
truncation to get to the minimal bosonic higher spin gauge theory
based on $hs(4)$ algebra as explained in
\refs\sezginone. Similarly, with an $SO(N)$ analogue, one would
end up with the proposal of Klebanov and Polyakov itself via
consistent trucation.

One could have started with more than one (say $M$) coincident
D5--branes in $AdS_5 \times S^5$. This should lead to a $U(M)$ valued
higher spin spectrum. Such higher spin theories have been considered
by Vasiliev \refs\vasilievfour.

It would be interesting to study the tensionless limit of open string
theory side explicitly based on the recent results
\refs{\tseytlin, \karchone, \dmw, \karchtwo}. Also, it may be 
possible to understand the higher spin spectrum considered here as
that of a `brane' in an appropriate higher spin gauge theory in
$AdS_5$. The spectrum on $AdS_4$ subspace may perhaps be thought of as
the goldstone and goldstino modes of broken higher spin symmetries of
a corresponding $AdS_5$ theory. Related ideas have been considered
recently in \refs\segal. We note that we only tested the duality that
we proposed here at the level of matching the spectra. However, the
way we `derived' it from a string theory context suggests that even
the interactions of the higher spin theory considered should be
reproducible from the dual modes. It will of course be interesting to
check this explicitly. One may also be able to define the bulk theory
in terms of the bilocal collective fields (see the footnote 4) as in
\refs\sumit. Finally, we have not considered the possible `double
trace' deformations of our free theory theory. The allowed set of
boundary conditions for the scalars in the bulk suggests the existence
of other (interacting) CFTs \refs\klebwit. It should be interesting to
explore that aspect as well. We hope to return to some of the open
questions in the future.
\bigskip

\noindent{\bf Acknowledgements:} 

I would like to thank David Berman, Michael Green, Sean Hartnoll,
Octavio Obregon, Giuseppe Policastro, Fernando Quevedo, Adam Ritz and
Aninda Sinha for discussions. I am particularly thankful to Sumit Das
for introducing me to the subject, collaboration and encouragement. I
thank Michael Green and Stefano Kovacs for their helpful comments on
the manuscript. My research is supported by a PPARC Research
Assistantship.

\appendix{A}{The conserved currents}
In this appendix we provide more detailed expressions of the conserved
currents listed in section 2. We closely follow
Refs. \refs{\anselmione, \anselmitwo, \vasilievfour, \vasilievfive,
\mikhailov}.

We seek conserved tensor primary current of the free field theory of a
complex scalar and a spinor in $d=3$ given in \bcft. More complete
expressions of the ones indicated in section 2 are as follows:
\eqn\qqcurrent{
\left({\cal J}^{(\bar q q)}_s\right)^\alpha = \sum_{k=0}^s
{(-1)^k (\del_{i_1} \cdots \del_{i_k} {\bar q}^m)
\sigma^\alpha_{mn} (\del_{i_{k+1}} \cdots \del_{i_s} q^n)\over 
\Gamma (k+1) \Gamma (s-k+1) \Gamma(k+ \half) \Gamma (s - k + \half)}
 - \rm traces,}
\eqn\psipsicurrent{
\left( {\cal J}^{(\bar \psi \psi)}_s \right)^\beta = \sum_{k=0}^{s-1} 
{ (-)^k (\del_{i_1} \cdots \del_{i_k} {\bar \psi}^i) \sigma^{\beta}_{ij}
\rho_{i_{k+1}} (\del_{i_{k+2}} \cdots \del_{i_s} \psi^j) \over 
\Gamma (k+1) \Gamma (s-k) \Gamma(k+ {3 \over 2}) \Gamma (s - k + \half)}
- \rm traces.}
In the above expressions $s = 0, 1, 2, \cdots$ and all the indices
$i_1, \cdots , i_s$ are understood to be completely
symmetrised. Further, as stated earlier, the indices $\alpha, \beta$ take
four values $\alpha, \beta = 0, 1, 2, 4$ and the corresponding
$\sigma^\alpha = \{ I, \sigma^i:~ i= 1,2,3\}$ with $I$ being the $2
\times 2$ identity matrix and $\sigma^i$ being the $SU(2)$ Pauli
matrices. For each spin $s$, we have a total of eight
currents. Similarly, the half-integer spin currents are given as
follows:
\eqn\qpsicurrent{
\eqalign{
\left( {\cal J}^{(\psi^{\star}q)}_s \right)^m_i &= \sum_{l,n,p
=0}^{\infty} {(-1)^l \Gamma ( p + l + n + {3 \over 2}) \over
\Gamma (l+1) \Gamma (n+1) \Gamma (p+1) 2^p \Gamma (p + l 
+ {3 \over 2}) \Gamma (p + n + {3 \over 2})} \cr 
& \Big[ (p + n + \half) \delta_{s-l-n-2p-\half} \eta_{i_1i_2} \cdots 
\eta_{i_{2p-1}i_{2p}} \del_{i_{2p+1}} \cdots \del_{i_{2p+l}}
\del^{j_1} \cdots \del^{j_p} \psi^{\star i} \cr
& \times \del_{i_{2p + l +1}} \cdots \del_{i_{p+l+n}} \del_{j_1}
\cdots \del_{j_p} q^m \cr
& -\half \delta_{s-l-n-2p -{3 \over 2}} \eta_{i_1i_2} \cdots 
\eta_{i_{2p-1}i_{2p}} \rho_{i_{2p+1}} \del_{i_{2p+2}} \cdots
\del_{i_{2p+l+1}} \del^{j_1}\cdots \del^{j_p} \rho_{i_{2p+l+2}}
\psi^{\star i} \cr
& \times \del_{i_{2p+l+3}} \cdots \del_{i_{2p+l+n+2}} \del_{j_1}
\cdots \del_{j_{p+1}} q^m \Big]
}}
and finally there is another set of currents obtained by $\psi^\star
\leftrightarrow \bar q$ and $q \leftrightarrow \psi$ in the expression
\qpsicurrent. Again there are $4+4 = 8$ currents with half
odd-integral spins as required for the model considered in section 2.

\listrefs
\end